\documentstyle[12pt]{article}

\newcommand{\eref}[1]{(\ref{#1})}

\newcommand{\cref}[1]{Chapter~\ref{#1}}
\newcommand{\beq}{\begin{equation}}
\newcommand{\eeq}{\end{equation}}
\newcommand{\ba}{\begin{array}}
\newcommand{\ea}{\end{array}}
\newcommand{\bcenter}{\begin{center}}
\newcommand{\ecenter}{\end{center}}

\def\IB{\relax\hbox{$\inbar\kern-.3em{\rm B}$}}
\def\IC{\relax\hbox{$\inbar\kern-.3em{\rm C}$}}
\def\ID{\relax\hbox{$\inbar\kern-.3em{\rm D}$}}
\def\IE{\relax\hbox{$\inbar\kern-.3em{\rm E}$}}
\def\IF{\relax\hbox{$\inbar\kern-.3em{\rm F}$}}
\def\IG{\relax\hbox{$\inbar\kern-.3em{\rm G}$}}
\def\IGa{\relax\hbox{${\rm I}\kern-.18em\Gamma$}}
\def\IH{\relax{\rm I\kern-.18em H}}
\def\IK{\relax{\rm I\kern-.18em K}}
\def\IL{\relax{\rm I\kern-.18em L}}
\def\IP{\relax{\rm I\kern-.18em P}}
\def\IR{\relax{\rm I\kern-.18em R}}
\def\IZ{\relax\ifmmode\mathchoice
{\hbox{\cmss Z\kern-.4em Z}}{\hbox{\cmss Z\kern-.4em Z}}
{\lower.9pt\hbox{\cmsss Z\kern-.4em Z}}
{\lower1.2pt\hbox{\cmsss Z\kern-.4em Z}}\else{\cmss Z\kern-.4em Z}\fi}
\def\II{\relax{\rm I\kern-.18em I}}

\def\sCC{{\kern 0.27em\vrule height1.45ex width0.03em depth0em
          \kern-0.30em\rm C}}
\def\C{{\mathchoice
  {\sCC}
  {\sCC}
  {\kern 0.225em \vrule height1.05ex width0.025em depth0em \kern-0.25em \rm C}
  {\kern 0.180em \vrule height0.78ex width0.02em depth0em \kern-0.2em \rm C}
        }}
\def\sHH{{\rm I\kern-.16em{}H}}
\def\H{{\mathchoice
  {\sHH}
  {\sHH}
  {\rm I\kern-.13em{}H}
  {\rm I\kern-.13em{}H} }}
\def\sNN{{\rm I\kern-.16em{}N}}
\def\N{{\mathchoice
  {\sNN}
  {\sNN}
  {\rm I\kern-.12em{}N}
  {\rm I\kern-.10em{}N} }}
\def\sPP{{\rm I\kern-.16em{}P}}
\def\P{{\mathchoice
  {\sPP}
  {\sPP}
  {\rm I\kern-.12em{}P}
  {\rm I\kern-.10em{}P} }}
\def\sQQ{{\kern 0.27em \vrule height1.45ex width0.03em depth0em
          \kern-0.30em \rm Q}}
\def\Q{{\mathchoice
        {\sQQ}
        {\sQQ}
  {\kern 0.225em \vrule height1.05ex width0.025em depth0em \kern-0.25em \rm Q}
  {\kern 0.180em \vrule height0.78ex width0.020em depth0em \kern-0.20em \rm Q}
        }}
\def\sRR{{\rm I\kern-0.16em{}R}}
\def\R{{\mathchoice
  {\sRR}
  {\sRR}
  {\rm I\kern-0.12em{}R}
  {\rm I\kern-0.10em{}R} }}
\def\sZZ{{\rm Z\kern-0.32em{}Z}}
\def\Z{{\mathchoice
  {\sZZ}
  {\sZZ} 
  {\rm Z\kern-0.3em{}Z}     
  {\rm Z\kern-0.25em{}Z} }}  
\def\ZZZ{{\rm Z\kern-0.24em{}Z}}
\def\sII{{\rm I\kern-0.16em{}I}}
\def\I{{\mathchoice
  {\sII}
  {\sII}
  {\rm I\kern-0.12em{}I}
  {\rm I\kern-0.10em{}I} }}

\def\dim{{\rm dim}}

\def\inbar{\,\vrule height1.5ex width.4pt depth0pt}
\font\cmss=cmss10 \font\cmsss=cmss10 at 7pt

\def\odd{\rm{odd}}
\def\even{\rm{even}}

\def\smiley{\hbox{\large$\bigcirc$\hspace{-0.80em}\raise.2ex
\hbox{$\cdot\cdot$}\kern-.61em\lower.2ex\hbox{\scriptsize$\smile$}}\ }
\def\frowny{\hbox{\large$\bigcirc$\hspace{-0.80em}\raise.2ex
\hbox{$\cdot\cdot$}\kern-.635em\lower.2ex\hbox{\scriptsize$\frown$}}\ }

\def\I{{\rlap{1} \hskip 1.6pt \hbox{1}}}

\newcommand{\gen}[1]{\langle #1 \rangle}

\makeatletter
\let\hangafter\@hangfrom
\makeatother

\renewcommand{\baselinestretch}{1.2}
\begin{document}
\renewcommand{\baselinestretch}{1}

\thispagestyle{empty}
{\flushright{\small MIT-CTP-3021\\hep-th/0009077\\}}
\vspace{.3in}
\begin{center}\LARGE {An Observation on Finite Groups and\\
	WZW Modular Invariants}
\end{center}

\vspace{.2in}
\begin{center}
{\large Bo Feng and Yang-Hui He\\}
\normalsize{fengb, yhe@ctp.mit.edu\footnote{
Research supported in part
by the CTP and the LNS of MIT and the U.S. Department of Energy 
under cooperative research agreement \# DE-FC02-94ER40818.}
\\}
\vspace{.2in} {\it Center for Theoretical Physics,\\ Massachusetts
Institute of Technology\\ Cambridge, Massachusetts 02139, U.S.A.\\}
\end{center}
\vspace{0.1in}

\begin{abstract}
In this short note, inspired by much recent activity centred around
attempts to formulate various correspondences between the
classification of affine $SU(k)$ WZW modular-invariant partition
functions and that of discrete finite subgroups of $SU(k)$, we present
a small and perhaps interesting observation in this light. 
In particular we show
how the groups generated by the permutation of the terms in the
exceptional $\widehat{SU(2)}$-WZW invariants encode the corresponding
exceptional $SU(2)$ subgroups.
\end{abstract}

\section{Introduction}
The ubiquitous $ADE$ meta-pattern of mathematics makes her mysterious emergence in
the classification of the modular invariant partition functions in Wess-Zumino-Witten
(WZW) models of rational conformal field theory (RCFT). Though this fact is by
now common knowledge, little is known about why {\it a fortiori} these invariants
should fall under such classification schemes \cite{CFT}.
Ever since the original work in the completion of the classification for $\widehat{su(2)}$
WZW invariants by Cappelli-Itzykson-Zuber \cite{su21,su22} as well as the subsequent
case for $\widehat{su(3)}$ by Gannon \cite{su31,su32}, many efforts have been
made to attempt to clarify the reasons behind the said emergence. These
include perspectives from lattice integrable systems where the invariants are
related to finite groups \cite{DiFrancesco}, and from generalised root systems and
$N$-colourability of graphs \cite{Reflection,Folding}. Furthermore, there has been a 
recent revival of interest in the matter as viewed from string theory
where sigma models and orbifold constructions are suggested to provide a link
\cite{HanHe,HeSong,Song}.

Let us first briefly review the situation at hand (much shall follow the conventions of
\cite{CFT} where a thorough treatment may be found). The $\widehat{g}_k$-WZW model
(i.e., associated to an affine Lie algebra $g$ at level $k$) is a non-linear
sigma model on the group manifold $G$ corresponding to the algebra $g$. Its
action is
\[
S^{\rm WZW} = \frac{k}{16\pi} \int_G \frac{d^2 x}{X_{\rm rep}} {\rm Tr}
	(\partial^\mu g^{-1} \partial_\mu g) + k \Gamma
\]
where $k \in \Z$ is called the level, $g(x)$, a matrix bosonic field
with target space\footnote{We are really integrating over the
	pull-back to the world sheet.}
 $G$ and $X_{\rm rep}$ the Dynkin index for the representation
of $g$. The first term is our familiar pull back in sigma models while the second
\[
\Gamma = \frac{-i}{24 \pi} \int_B \frac{d^3 y}{X_{\rm rep}} 
	\epsilon_{\alpha \beta \gamma} {\rm Tr}
	(\tilde{g}^{-1} \partial^\alpha\tilde{g} \tilde{g}^{-1} 
	\partial^\beta\tilde{g} \tilde{g}^{-1} \partial^\gamma\tilde{g} )
\]
is the WZW term added to ensure conformal symmetry. $B$ is a manifold such that
$\partial B = G$ and $\tilde{g}$ is the subsequent embedding of $g$ into $B$.
The conserved currents $J(z) := \sum\limits_a J^a t^a$ and
$J^a := \sum\limits_{n \in \Z} J^a_n z^{-n-1}$ 
(together with an independent anti-holomorphic copy)
form a {\bf current algebra} which is precisely the level $k$ affine algebra 
$\widehat{g}$:
\[
\left[ J^a_n, J^b_m \right] = i \sum\limits_c f_{abc} J^c_{n+m} + 
	k n \delta_{a b} \delta_{n+m,0}.
\]
The energy momentum tensor $T(z) = \frac{1}{d+k} \sum\limits_a J^a J^a$
with $d$ the dual Coxeter number of $g$ furnishes a Virasoro algebra
with central charge
\[
c(\widehat{g}_k) = \frac{k \dim g}{k+d}.
\]

Moreover, the primary fields are in 1-1 correspondence with the heighest weights
$\widehat{\lambda} \in P^k_+$ of $\widehat{g}$, which, being of a finite number,
constrains the number of primaries to be finite, thereby making WZW a RCFT.
The {\bf fusion algebra} of the primaries $\phi$ for this RCFT is consequently given by
$\phi_i \times \phi_j = \sum\limits_{\phi_k^*} 
{\cal N}_{\phi_i \phi_j}^{\phi_k^*}\phi_k^*$, or in the integrable representation
language of the affine algebra:
\[
\widehat{\lambda} \otimes \widehat{\mu} =
\bigoplus_{\widehat{\nu} \in P_+^k} 
{\cal N}_{\widehat{\lambda}\widehat{\mu}}^{\widehat{\nu}}\ \widehat{\nu}.
\]

The Hilbert Space of states decomposes into holomorphic and anti-holomorphic parts
as
${\cal H} = \bigoplus\limits_{\widehat{\lambda},\widehat{\xi} \in P^{(k)}_+}
{\cal M}_{\widehat{\lambda},\widehat{\xi}} H_{\widehat{\lambda}} 
\otimes H_{\widehat{\xi}}$ 
with the {\bf mass matrix} 
${\cal M}_{\widehat{\lambda},\widehat{\xi}}$
counting the multiplicity
of the $H$-modules in the decomposition. Subsequently, the partition function
over the torus, 
$Z(q) := {\rm Tr}_{\cal H} q^{L_0 - \frac{c}{24}} 
\bar{q}^{\bar{L}_0 - \frac{c}{24}}$
with $q := e^{2 \pi i \tau}$ reduces to
\beq
\label{Z}
Z(\tau) = \sum_{\widehat{\lambda},\widehat{\xi} \in P^k_+}
        \chi_{\widehat{\lambda}}(\tau)
        {\cal M}_{\widehat{\lambda},\widehat{\xi}}
        \bar{\chi}_{\widehat{\xi}}(\bar{\tau})
\eeq
with $\chi$ being the affine characters of $\widehat{g}_k$.
Being a partition function on the torus, (\ref{Z}) must obey the $SL(2;\Z)$
symmetry of $T^2$, i.e., it must be invariant under the {\bf modular group}
generated by $S : \tau \rightarrow -1/\tau$ and $T : \tau \rightarrow \tau + 1$.
Recalling the modular transformation properties of the affine characters, viz.,
\[
\begin{array}{lll}
T : \chi_{\widehat{\lambda}} & \rightarrow & \sum\limits_{\widehat{\mu} \in P^k_+}
	{\cal T}_{\widehat{\lambda}\widehat{\mu}} \chi_{\widehat{\mu}} \\
S : \chi_{\widehat{\lambda}} & \rightarrow & \sum\limits_{\widehat{\mu} \in P^k_+}
	{\cal S}_{\widehat{\lambda}\widehat{\mu}} \chi_{\widehat{\mu}}
\end{array}
\]
with
\[
\begin{array}{lll}
{\cal T}_{\widehat{\lambda} \widehat{\mu}} & = & \delta_{\widehat{\lambda}\widehat{\mu}}
	e^{\pi i (\frac{| \widehat{\lambda} + \widehat{\rho} |^2}{k+d} - 
			\frac{|\widehat{\rho}|^2}{d})} \\
{\cal S}_{\widehat{\lambda} \widehat{\mu}} & = & K\sum\limits_{w \in W} \epsilon(w)
	e^{-\frac{2 \pi i}{k + d} (w(\lambda + \rho), \mu + \rho)}
\end{array}
\]
where $\widehat{\rho}$ is the sum of the fundamental weights, $W$, the Weyl group
and $K$, some proportionality constant.
Modular invariance of (\ref{Z}) then implies $\left[{\cal M}, {\cal S} \right] =
\left[{\cal M}, {\cal T} \right] = 0$. 
The problem of classfication of the physical
modular invariants of $\widehat{g}_k$-WZW then amounts to solving for
{\em all nonnegative
integer matrices ${\cal M}$ such that ${\cal M}_{00} = 1$} (so as to guarantee uniqueness
of vacuum) and {\em satisfying these commutant relations.}

The fusion coefficients ${\cal N}$ can be, as it is with modular tensor categories
(q.v. e.g. \cite{HeSong}), related to the matrix ${\cal S}$ by the celebrated
Verlinde Formula:
\beq
\label{Verlinde}
{\cal N}^t_{rs} = \sum\limits_m 
	\frac{{\cal S}_{rm}{\cal S}_{sm}{\cal S}^{-1}_{mt}}{{\cal S}_{0m}}.
\eeq
Furthermore, in light of the famous McKay Correspondence (Cf. e.g. \cite{HanHe,HeSong}
for discussions of the said correspondence in this context), to establish 
correlations between modular invariants and graph theory, one can
chose a fundamental representation $f$ and regard $(N)_{st} := {\cal N}^t_{fs}$ as
an adjacency matrix of a finite graph. Conversely out of the adjacency
matrix $(G)_{st}$ for some finite graph, one can extract a set of matrices 
$\{(N)_{st}\}_i$ such that $N_0 = \I$ and $N_f = G$. We diagonalise $G$ as
${\cal S} \Delta {\cal S}^{-1}$ and define, as inspired by (\ref{Verlinde}),
the set of matrices 
$N_r := \{(N)_{st}\}_r = \sum\limits_m 
\frac{{\cal S}_{rm}{\cal S}_{sm}{\cal S}^{-1}_{mt}}{{\cal S}_{0m}}$, which clearly
satisfy the constriants on $N_{0,f}$.
This set of matrices $\{N_i\}$, each associated to a vertex in the judiciously chosen graph,
give rise to a {\bf graph algebra} and appropriate subalgebras thereof, by virtue of
matrix multiplication, constitute a representation for the fusion algebra, i.e.,
$N_i \cdot N_j = \sum\limits_k {\cal N}_{ij}^k N_k$. In a more axiomatic language,
the Verlinde equation (\ref{Verlinde}) is essentially the inversion of the McKay
composition
\beq
\label{McKay}
R_r \otimes R_s = \bigoplus\limits_t {\cal N}^t_{rs} R_t
\eeq
of objects $\{R_i\}$ in a (modular) tensor category. 
The ${\cal S}$ matrices are then the characters
of these objects and hence the matrix of eigenvectors of $G = {\cal N}^t_{rs}$ once
fixing some $r$ by definition (\ref{McKay}). The graph algebra is essentially the
set of these matrices ${\cal N}^t_{rs}$ as we extrapolate $r$ from 0 (giving $\I$)
to some fixed value giving the graph adjacency matrix $G$.

Thus concludes our brief review on the current affair of things. Let
us now proceed to present our small observation.
\section*{Nomenclature}
Throughout the paper, unless otherwise stated, we shall adhere to the
folloing conventions:
$G_n$ is group $G$ of order $n$.
$\gen{x_i}$ is the group generated by the (matrix) elements $\{x_i\}$.
$k$ is the level of the WZW modular invariant partition function $Z$.
$\chi$ is the affine character of the algebra $\widehat{g}$.
${\cal S},{\cal T}$ are the generators of the modular group
$SL(2;\Z)$ whereas $S,T$ will be these matrices in a new basis, to be
used to generate a finite group. $E_{6,7,8}$ are the ordinary
tetrahedral, octahedral and icosahedral groups while
$\widehat{E_{6,7,8}}$ are their binary counterparts. Calligraphic font
(${\cal A,D,E}$) shall be reserved for the names of the modular invariants.

\section{$\widehat{su(2)}$-WZW}
The modular invariants of $\widehat{su(2)}$-WZW were originally classified in
the celebrated works of \cite{su21,su22}.
The only solutions of the abovementioned conditions for 
$k,{\cal S}, {\cal T}$ and ${\cal M}$ give rise to the following:
\beq
\label{su2ST}
{\cal S}_{ab}=\sqrt{{2\over k+2}}\,\sin(\pi\,{(a+1)(b+1)\over {k+2}}),\qquad 
{\cal T}_{ab}=\exp[ \pi i ({(a+1)^2\over 2(k+2)} - {1 \over 4})]\,\delta_{a,b}
\qquad a,b = 0,...,k
\eeq
with the partition functions
\beq
\label{su2Z}
\begin{array}{lll}
k & {\cal A}_{k+1} & Z = \sum\limits_{n=0}^{k} |\chi_n|^2 \\

k = 4m & {\cal D}_{2m+2} & Z = \sum\limits_{n=0,\even}^{2m-2} 
	|\chi_n + \chi_{k - n}|^2 + 2 |\chi_{2m}|^2 \\

k = 4m-2 & {\cal D}_{2m+1} & Z = |\chi_{k \over 2}|^2 + 
	\sum\limits_{n=0,\even}^{4m-2} |\chi_n|^2 + 
	\sum\limits_{n=1,\odd}^{2m-1}(\chi_n \bar{\chi}_{k-n} + c.c.) \\

k = 10 & {\cal E}_6 & Z = |\chi_0+\chi_6|^2+|\chi_3+\chi_7|^2+|\chi_4+\chi_{10}|^2 \\

k = 16 & {\cal E}_7 & Z = |\chi_0+\chi_{16}|^2+|\chi_4+\chi_{12}|^2+|\chi_6+\chi_{10}|^2
	+(\bar{\chi}_8(\chi_2+\chi_{14}) + c.c.) \\

k = 28 & {\cal E}_8 & Z = |\chi_0+\chi_{10}+\chi_{18}+\chi_{28}|^2+
	|\chi_6+\chi_{12}+\chi_{16}+\chi_{22}|^2 \\
\end{array}
\eeq

We know of course that the simply-laced simple Lie algebras, as well as the
discrete subgroups of $SU(2)$ fall precisely under such a classification. The now
standard method is to associate the modular invariants to subalgebras of the
graph algebras constructed out of the respective ADE-Dynkin Diagram. This is done in
the sense that the adjacency matrices of these diagrams\footnote{These are the well-known 
	symmetric matrices of eigenvalues $\le 2$, or equivalently, the McKay 
	matrices for $SU(2)$; for a discussion on this point q.v. e.g. \cite{He}.}
are to define $N_1$ and subsets of $N_i$ determine the fusion rules. The correspondence
is rather weak, for in addition to the necessity of the truncation to
subalgebras, only $A_k$, $D_{2k}$
and $E_{6,8}$ have been thus related to the graphs while $D_{2k+1}$ and $E_7$
give rise to negative entries in ${\cal N}_{ij}^k$. However as an encoding process,
the above correspondences has been very efficient, especially in generalising to WZW
of other algebras.

The first attempt to explain the ADE scheme in the $\widehat{su(2)}$ modular invariants
was certainly not in the sophistry of the above context. It was in fact done in the
original work of \cite{su22}, where the authors sought to relate their invariants
to the discrete subgroups of $SO(3) \cong SU(2)/\Z_2$. It is under the inspiration of
this idea, though initially abandoned ({\it cit. ibid.}), that the current
writing has its birth. We do not promise to find a stronger correspondence, yet we
shall raise some observations of interest.

The basic idea is simple. To ourselves we pose the obvious question: what, algebraically
does it mean for our partition functions (\ref{su2Z}) to be modular invariant?
It signifies that the action by ${\cal S}$ and ${\cal T}$ thereupon must permute the
terms thereof in such a way so as not to, by virtue of the transformation properties
of the characters (typically theta-functions), introduce extraneous terms.
In the end of the monumental work \cite{su22}, the authors, as a diversion,
used complicated identities of theta and eta functions to rewrite the
${\cal E}_{6,7,8}$ cases of (\ref{su2Z}) into sum of
terms on whose powers certain combinations of ${\cal S}$ and ${\cal T}$ act. These
combinations were then used to generate finite groups which in the
case of ${\cal E}_6$, did
give the ordinary tetrahedral group $E_6$ and ${\cal E}_8$, the
ordinary icosahedral group $E_8$, which are
indeed the finite groups associated to these Lie algebras, a fact which dates back to
F. Klein. As a postlude, \cite{su22} then speculated upon the reasons for this correspondence
between modular invariants and these finite groups, as being attributable to the
representation of the modular groups over finite fields, since afterall
$E_6 \cong PSL(2;\Z_3)$ and $E_8 \cong PSL(2;\Z_4) \cong PSL(2;\Z_5)$.

We shall not take recourse to the complexity of manipulation of theta functions
and shall adhere to a pure group theoretic perspective. We translate the aforementioned
concept of the permutation of terms into a vector space language.
First we interpret the characters appearing in (\ref{su2Z}) as basis upon which
${\cal S}$ and ${\cal T}$ act. For the $k$-th level they are defined as the canonical
bases for $\C^{k+1}$:
\[
\chi_0 := (1,0,...,0); \quad ... \quad \chi_i := (\I)_{i+1}; 
\quad ... \quad \chi_k := (0,0,...,1).
\]
Now ${\cal T}$ being diagonal clearly maps these vectors to multiples of themselves
(which after squaring the modulus remain uneffected); the interesting
permutations are performed by ${\cal S}$.
\subsection{The $E_6$ Invariant}
Let us first turn to the illustrative example of ${\cal E}_6$. 
From $Z$ in (\ref{su2Z}), we see that we are clearly interested in the vectors
$v_1 := \chi_0 + \chi_6 = (1, 0, 0, 0, 0, 0, 1, 0, 0, 0, 0),
 v_2 := \chi_4 + \chi_{10} = (0, 0, 0, 0, 1, 0, 0, 0, 0, 0, 1)$ and
$v_3 := \chi_3 + \chi_7 = (0, 0, 0, 1, 0, 0, 0, 1, 0, 0, 0)$.
Hence (\ref{su2ST}) gives
${\cal T} : v_1 \rightarrow e^{{-{5 \pi i} \over {24}} } v_1$,
${\cal T} : v_2 \rightarrow e^{{{19 \pi i} \over {24}} } v_2$ and
${\cal T} : v_3 \rightarrow e^{{{ 5 \pi i} \over {12}} } v_3$.
Or, in other words in the subspace spanned by $v_{1,2,3}$, ${\cal T}$ acts
as the matrix $T := {\rm Diag}(e^{{-{5 \pi i} \over {24}}},
	e^{{{19 \pi i} \over {24}}}, e^{{{ 5 \pi i} \over {12}}})$.
Likewise, ${\cal S}$ becomes a 3 by 3 matrix; we present them below:
\beq
\label{e6newST}
S =
\left( 
\matrix{ {1\over 2} & {1\over 2} & {1\over {{\sqrt{2}}}} \cr
    {1\over 2} & {1\over 2} & -{1\over {{\sqrt{2}}}} \cr
    {1\over {{\sqrt{2}}}} & -{1\over {{\sqrt{2}}}} & 0 }
\right)
\qquad
T =
\left(
\matrix{{e^{-{{5 \pi i}\over {24}}}} & 0 & 0
   \cr 0 & {e^{{{19 \pi i}\over {24}}}} & 0 ,
   \cr 0 & 0 & {e^{{{5 \pi i}\over {12}}}}}
\right)
\eeq

Indeed no extraneous vectors are involved, i.e., of the 11 vectors  $\chi_i$ 
and all combinations of sums thereof, only the combinations $v_{1,2,3}$ appear
after actions by ${\cal S}$ and ${\cal T}$. This closure of course is what is needed
for modular invariance. What is worth of note, is that we have collapsed an
11-dimensional representation of the modular group acting on $\{\chi_i\}$, to
a (non-faithful) 3-dimensional representation which corresponds the
subspace of interest (of the initial $\C^{11}$) by virtue of the appearance of the
terms in the associated modular invariant. Moreover the new matrices
$S$ and $T$, being of finite order (i.e., $\exists m,n \in \Z_+$ s.t. $S^m = T^n = \I$),
actually generate a {\it finite group}. It is this finite group that we shall compare to
the ADE-subgroups of $SU(2)$.

The issue of the finiteness of the initial group generated 
by ${\cal S}$ and ${\cal T}$ was addressed in a recent work by Coste and Gannon \cite{Coste}.
Specifically, the group
\beq
\label{poly}
P := \{S,T | T^N = S^2 = (ST)^3 = \I \},
\eeq
generically known as the {\em polyhedral (2,3,N) group}, is infinite
for $N > 5$.
On the other hand, for $N=2,3,4,5$, $G \cong \Gamma / \Gamma(N) := SL(2;\Z/N\Z)$,
which, interestingly enough, for these small values are, the symmetric-3, the tetrahedral,
the octahedral and icosahedral groups respectively.

We see of course that our matrices in \eref{e6newST} satisfy
the relations of \eref{poly} with $N=48$ (along with additional
relations of course) and hence generates a subgroup of $P$. Indeed,
$P$ is the modular group in a field of finite characteristic $N$ and
since we are dealing with nonfaithful representations of the modular
group, the groups generated by $S,T$, as we shall later see, in the
cases of other modular invariants are all finite subgroups of $P$.

In our present case, $G=\gen{S,T}$ is of order 1152. Though $G$ itself
may seem unenlightening, upon closer inspection we find that it has
12 normal subgroups $H \lhd G$ and {\em only one} of which is of
order 48. In fact this $H_{48}$ is $\IZ_4 \times \IZ_4 \times
\IZ_3$. The observation is that the quotient group formed between $G$
and $H$ is precisely the binary tetrahedral group $\widehat{E_6}$, i.e.,
\beq
\label{e6res}
G_{1152} / H_{48} \cong \widehat{E_6}.
\eeq

We emphasize again the {\em uniqueness} of this procedure: as will be
with later examples, given $G({\cal E}_6)$, there exists a unique
normal subgroup which can be quotiented to give $\widehat{E_6}$, and
moreover there does not exist a normal subgroup which could be used to
generate the other exceptional groups, viz.,  $\widehat{E_{7,8}}$. We
shall later see that such a 1-1 correpondence between the exceptional modular
invariants and the exceptional discrete groups persists.

This is a pleasant surprise; it dictates that the symmetry group
generated by the permutation of the terms in the ${\cal E}_6$ modular
invariant partition function of $\widetilde{SU(2)}$-WZW, upon
appropriate identification, is exactly the symmetry group assocaited
to the $\widehat{E_6}$ discrete subgroup of $SU(2)$. Such a correspondence may
{\it a priori} seem rather unexpected.
\subsection{Other Invariants}
It is natural to ask whether similar circumstances arise for the
remaining invariants. Let us move first to the the case of $E_8$. By
procedures completely analogous to \eref{e6newST} as applied to the
partition function in \eref{su2Z}, we see that the basis is composed
of $v_1 =
\chi_0 + \chi_{10} + \chi_{18} + \chi_{28} = \{1, 0, 0, 0, 0, 0, 0, 0,
0, 0, 1, 0, 0, 0, 0, 0, 0, 0, 1, 0, 0, 0, 0, 0, 0, 0, 0, 0, 1\}$ and
$v_2 = \chi_6 + \chi_{12} + \chi_{16} + \chi_{22} = \{0, 0, 0, 0, 0,
0, 1, 0, 0, 0, 0, 0, 1, 0, 0, 0, 1, 0, 0, 0, 0, 0, 1, 0, 0, 0, 0, 0, 0
\}$, under which $S$ and $T$ assume the forms as summarised in
Table \ref{Esummary}.

This time $G = \gen{S,T}$ is of order 720, with one {\it unique}
normal subgroup of order 6 (in fact $\IZ_6$). Moreover we find that
\beq\label{e8res}
G_{720} / H_6 \cong \widehat{E_8},
\eeq
in complete analogy with \eref{e6res}. Thus once again, the symmetry
due to the permutation of the terms inherently encode the associated
discrete $SU(2)$ subgroup.

What about the remaining exceptional invariant, $E_7$? The basis as
well as the matrix forms of $S,T$ thereunder are again presented in
Table \ref{Esummary}. The group generated thereby is of order 324,
with 2 non-trivial normal subgroups of orders 27 and
108. Unfortunately, no direct quotienting could possibly give the
binary octahedral group here. However $G / H_{27}$ gives a group of
order 12 which is in fact the {\it ordinary} octahedral group $E_7=A_4$, which
is in turn isomorphic to $\widehat{E_7} / \IZ_2$. Therefore for our present
case the situation is a little more involved:
\beq\label{e7res}
G_{324} / H_{27} \cong \widehat{E_7} / \IZ_2 \cong E_7.
\eeq

We recall \cite{CFT} that a graph algebra \eref{Verlinde} based on the
Dynkin diaram of $E_7$ has actually not been succesully constructed
for the $E_7$ modular invariant. Could we speculate that the slight
complication of \eref{e7res} in comparison with \eref{e6res} and
\eref{e8res} be related to this failure?

\beq
\label{Esummary}
{\scriptsize \hspace{-0.5cm}
\begin{array}{|c|c|c|}
\hline
	& \mbox{\normalsize Matrix Generators}	&  \mbox{\normalsize Basis} \\ \hline
E_6	& S = \left( 
	\matrix{ {1\over 2} & {1\over 2} & {1\over {{\sqrt{2}}}} \cr
    		{1\over 2} & {1\over 2} & -{1\over {{\sqrt{2}}}} \cr
    		{1\over {{\sqrt{2}}}} & -{1\over {{\sqrt{2}}}} & 0 }
	\right)
	  T = \left(\matrix{{e^{{{-5 \pi i}\over {24}}}} & 0 & 0
   		\cr 0 & {e^{{{19 \pi i}\over {24}}}} & 0 ,
   		\cr 0 & 0 & {e^{{{5 \pi i}\over {12}}}}}
	\right)
	&
	\begin{array}{l}
		v_1 = \chi_0 + \chi_6 \\
		v_2 = \chi_4 + \chi_{10} \\
		v_3 = \chi_3 + \chi_7 \\
	\end{array}
	\\
\hline
E_7	& 
	\begin{array}{c}
	S = {1 \over 3} \left(
 	\matrix{ \sin ({{\pi }\over {18}}) + \sin ({{17\,\pi }\over {18}}) & 
   		\sin ({{5\,\pi }\over {18}}) + \sin ({{85\,\pi }\over {18}}) & 
   		\sin ({{7\,\pi }\over {18}}) + \sin ({{119\,\pi }\over {18}}) & 2 & 1 \cr 
   		\sin ({{5\,\pi }\over {18}}) + \sin ({{13\,\pi }\over {18}}) & 
   		\sin ({{25\,\pi }\over {18}}) + \sin ({{65\,\pi }\over {18}}) & 
   		\sin ({{35\,\pi }\over {18}}) + \sin ({{91\,\pi }\over {18}}) & 2 & 1 \cr 
   		\sin ({{7\,\pi }\over {18}}) + \sin ({{11\,\pi }\over {18}}) & 
   		\sin ({{35\,\pi }\over {18}}) + \sin ({{55\,\pi }\over {18}}) & 
   		\sin ({{49\,\pi }\over {18}}) + \sin ({{77\,\pi }\over {18}}) & -2 & -1
    		\cr 1 & 1 & -1 & 1 & -1 \cr 1 & 1 & -1 & -2 & 2 \cr  } 
	\right) \\
	T = \left(\matrix{ {e^{{{-2\,i}\over 9}\,\pi }} & 0 & 0 & 0 & 0 \cr 0 & 
	   	{e^{{{4\,i}\over 9}\,\pi }} & 0 & 0 & 0 \cr 0 & 0 & 
   		{e^{{{-8\,i}\over 9}\,\pi }} & 0 & 0 \cr 0 & 0 & 0 & 1 & 0 \cr 0 & 0 & 0 & 
   		0 & 1 \cr  }
	\right)
	\end{array}
	&
	\begin{array}{l}
		v_1 = \chi_0+\chi_{16} \\
		v_2 = \chi_4+\chi_{12} \\
		v_3 = \chi_6+\chi_{10} \\
		v_4 = \chi_8 \\
		v_5 = \chi_2+\chi_{14} \\
	\end{array} \\
\hline
E_8	&
	\begin{array}{c}
	S = {1 \over \sqrt{15}} \left(
	\matrix{ \sin ({{\pi }\over {30}}) + \sin ({{11\,\pi }\over {30}}) + 
    \sin ({{19\,\pi }\over {30}}) + \sin ({{29\,\pi }\over {30}}) & 
   \sin ({{7\,\pi }\over {30}}) + \sin ({{77\,\pi }\over {30}}) + 
    \sin ({{133\,\pi }\over {30}}) + \sin ({{203\,\pi }\over {30}}) \cr 
   \sin ({{7\,\pi }\over {30}}) + \sin ({{13\,\pi }\over {30}}) + 
    \sin ({{17\,\pi }\over {30}}) + \sin ({{23\,\pi }\over {30}}) & 
   \sin ({{49\,\pi }\over {30}}) + \sin ({{91\,\pi }\over {30}}) + 
    \sin ({{119\,\pi }\over {30}}) + \sin ({{161\,\pi }\over {30}}) \cr  } \right)\\
	T = \left(\matrix{ {e^{{{-7\,i}\over {30}}\,\pi }} & 0 \cr 0 & 
   	{e^{{{17\,i}\over {30}}\,\pi }} \cr  } \right)
	\end{array}
	&
	\begin{array}{l}
		v_1 = \chi_0+\chi_{10}+\chi_{18}+\chi_{28} \\
		v_2 = \chi_6+\chi_{12}+\chi_{16}+\chi_{22} \\
	\end{array} \\
\hline
\end{array}
}
\eeq

We shall pause here with the exceptional series as for the infinite
series the quotient of the polyhedral $(2,3,N)$ will never give any
abelian group other than $\IZ_{1,2,3,4,6}$ or any dihedral group other
than $D_{1,3}$ \cite{Com}. More complicated procedures are called for
which are yet to be ascertained \cite{Prog}, though we remark here
briefly that for the $A_{k+1}$ series, since $Z$
is what is known as the {\it diagonal invariant}, i.e., it includes all
possible $\chi_n$-bases, we need not perform any basis change and
whence $S,T$ are simply the original ${\cal S,T}$ and there is an
obvious relationship that $G := \gen{T^8} \cong \IZ_{k+2}
:= A_{k+1}$.

Incidentally, we can ask ourselves whether any such correspondences
could possibly hold for the {\em ordinary} exceptional groups. From
\eref{e7res} we see that $G({\cal E}_7)/H_{27}$ does indeed correspond
to the ordinary octahedral group. Upon further investigation, we find
that $G({\cal E}_6)$ could not be quotiented to give the ordinary $E_6$
while $G({\cal E}_8)$ does have a normal subgroup of order 12 which
could be quotiented to give the ordinary $E_8$.
Without much further
ado for now, let us summarise these results:
\[
\ba{|c|c|c|c|c|}
\hline
	& G := \gen{S,T} & \mbox{Normal Subgroups} &
		\multicolumn{2}{|c|}{\mbox{Relations}} \\ \hline
{\cal E}_6 & G_{1152} & H_{3,4,12,16,48,64,192,192',384,576} &
	G_{1152} / H_{48} \cong \widehat{E_6} & - \\
{\cal E}_7 & G_{324} & H_{27,108} & G_{324} / H_{27} \cong \widehat{E_7} / \IZ_2
	& G_{324} / H_{27} \cong E_7 \\
{\cal E}_8 & G_{720} & H_{2,3,4,6,12,120,240,360} & G_{720} / H_6
	\cong \widehat{E_8} & G_{720} / H_{12} \cong E_8 \\
\hline
\ea
\]
\section{Prospects: $\widehat{su(3)}$-WZW and Beyond?}
There has been some recent activity \cite{DiFrancesco,HanHe,HeSong,Song}
in attempting to explain the patterns emerging in the modular invariants
beyond $\widehat{su(2)}$. Whether from the perspective of integrable systems,
string orbifolds or non-linear sigma models, proposals of the invariants
being related to subgroups of $SU(n)$ have been made.
It is natural therefore for us to inquire whether the correspondences from the
previous subsection between $\widehat{su(n)}$-WZW and the discrete subgroups of
$SU(n)$ for $n=2$ extend to $n=3$.

We recall from \cite{su31,su32} that the modular invariant partition functions
for $\widehat{su(3)}$-WZW have been classified to be the following:
\beq
\label{su3Z}
\begin{array}{lll}
{\cal A}_k & := &  \sum\limits_{\lambda \in P^k} |\chi_\lambda^k|^2,\qquad
	\forall k\ge 1; \cr

{\cal D}_k & := & \sum\limits_{(m,n)\in P^k} \chi^k_{m,n} \chi^{k*}_{\omega^{k(m-n)}(m,n)}, 
	\quad {\rm for}~k \not\equiv 0 \bmod 3~{\rm and}~k \ge 4; \cr

{\cal D}_k & := & {1\over 3}
	\sum\limits_{{(m,n)\in P^k} \atop {m \equiv n \bmod 3}}
	|\chi^k_{m,n}+\chi_{\omega(m,n)}^k+\chi^k_{\omega^2(m,n)}|^2; \cr

{\cal E}_5 & := & |\chi^5_{1,1}+\chi^5_{3,3}|^2+|\chi^5_{1,3}+\chi^5_{4,3}|^2+
	|\chi^5_{3,1}+\chi^5_{3,4}|^2 + \\&&
	|\chi^5_{3,2}+\chi^5_{1,6}|^2+|\chi^5_{4,1}+
	\chi^5_{1,4}|^2+|\chi^5_{2,3}+\chi^5_{6,1}|^2; \cr

{\cal E}_9^{(1)} & := & |\chi^9_{1,1}+\chi^9_{1,10}+\chi^9_{10,1}+\chi^9_{5,5}+
	\chi^9_{5,2}+\chi^9_{2,5}|^2+2|\chi^9_{3,3}+\chi^9_{3,6}+\chi^9_{6,3}|^2; \cr 

{\cal E}_9^{(2)} & := &  |\chi^9_{1,1}+\chi^9_{10,1}+\chi^9_{1,10}|^2+|\chi^9_{3,3}+
	\chi^9_{3,6}+\chi^9_{6,3}|^2+ 2|\chi^9_{4,4}|^2 \\&&
	+ |\chi^9_{1,4}+\chi^9_{7,1}+ \chi^9_{4,7}|^2+|\chi^9_{4,1}+\chi^9_{1,7}+
	\chi^9_{7,4}|^2+|\chi^9_{5,5}+\chi^9_{5,2}+\chi^9_{2,5}|^2 \\&&
	+ (\chi^9_{2,2}+\chi^9_{2,8}+\chi^9_{8,2})\chi^{9*}_{4,4}+
	\chi^9_{4,4}(\chi^{9*}_{2,2}+\chi^{9*}_{2,8}+\chi^{9*}_{8,2}); \cr

{\cal E}_{21} & := & |\chi^{21}_{1,1}+\chi^{21}_{5,5}+\chi^{21}_{7,7}+
	\chi^{21}_{11,11}+\chi^{21}_{22,1}+\chi^{21}_{1,22}
	+\chi^{21}_{14,5}+\chi^{21}_{5,14}+\chi^{21}_{11,2}+\chi^{21}_{2,11}+\chi^{21}_{10,7}
	+\chi^{21}_{7,10}|^2 \\&&
	+|\chi^{21}_{16,7}+\chi^{21}_{7,16}+\chi^{21}_{16,1}+\chi^{21}_{1,16}+
		\chi^{21}_{11,8}+\chi^{21}_{8,11}
	+\chi^{21}_{11,5}+\chi^{21}_{5,11}+
		\chi^{21}_{8,5}+\chi^{21}_{5,8}+\chi^{21}_{7,1}+\chi^{21}_{1,7}|^2; \cr
\end{array}
\eeq
where we have labeled the level $k$ explicitly as subscripts. Here the highest weights
are labeled by two integers $\lambda = (m,n)$ as in the set
\[	
P^k := \{\lambda = m \beta_1 + n\beta_2\,|\,m,n\in\Z,~ 0 <  m, n, m+n < k+3 \}
\]
and $\omega$ is the operator $\omega :(m,n) \rightarrow (k+3-m-n,n)$. The modular
matrices are simplified to
\beq
\label{su3ST}
\begin{array}{lll}
{\cal S}_{\lambda \lambda'} & = & {-i \over {\sqrt{3}(k+3)}} \bigr\{ e_k(2mm'+mn'+nm'+2nn')
	+e_k(-mm'-2mn'-nn'+nm') \\&& +e_k(-mm'+mn'-2nm'-nn')-e_k(-2mn'-mm'-nn'-2nm') \\&&
	-e_k(2mm'+mn'+nm'-nn')-e_k(-mm'+mn'+nm'+2nn')\} \cr
{\cal T}_{\lambda \lambda'} & = & e_k(-m^2-mn-n^2+k+ 3)\,\delta_{m,m'}\,\delta_{n,n'}
\end{array}
\eeq
with $e_k(x) := \exp[{-2\pi i x\over 3(k+3)}]$.

We imitate the above section and attempt to generate various finite
groups by $S,T$ under appropriate transformations from \eref{su3ST} to
new bases. We summarise the results below:
\[
\ba{|c|c|c|}
\hline
& {\rm Basis} & G:=\gen{S,T}\\
\hline
{\cal E}_5 & \{\chi_{1,1}+\chi_{3,3}; \chi_{1,3}+\chi_{4,3};
         \chi_{3,1}+\chi_{3,4}; \chi_{3,2}+\chi_{1,6};
         \chi_{4,1}+\chi_{1,4}; \chi_{2,3}+\chi_{6,1}\}
& G_{1152}\\ \hline
{\cal E}_9^{(1)} &
\{\chi_{1,1}+\chi_{1,10}+\chi_{10,1}+\chi_{5,5}+\chi_{5,2}+\chi_{2,5}];
  \chi_{3,3}+\chi_{3,6}+\chi_{6,3}\}
& G_{48}\\ \hline
{\cal E}_9^{(2)} & 
\ba{c}
\{\chi_{1,1}+\chi_{1,10}+\chi_{10,1};
\chi_{5,5}+\chi_{5,2}+\chi_{2,5}; \\
\chi_{3,3}+\chi_{3,6}+\chi_{6,3}; \chi_{4,4};
\chi_{4,1}+\chi_{1,7}+\chi_{7,4}; \\
\chi_{1,4}+\chi_{7,1}+\chi_{4,7};
\chi_{2,2}+\chi_{2,8}+\chi_{8,2}\}
\ea
& G_{1152}\\ \hline
{\cal E}_{21} &
\ba{c}
\{\chi_{1,1}+\chi_{5,5}+\chi_{7,7}+\chi_{11,11}+\chi_{22,1}+
 \chi_{1,22}+\chi_{14,5}+\chi_{5,14}+ \\
 \chi_{11,2}+\chi_{2,11}+ \chi_{10,7}+\chi_{7,10}; \\
 \chi_{16,7}+\chi_{7,16}+\chi_{16,1}+\chi_{1,16}+\chi_{11,8}+
 \chi_{8,11} +\\
 \chi_{11,5}+\chi_{5,11}+\chi_{8,5}+\chi_{5,8}+
 \chi_{1,7}+\chi_{7,1}\}
\ea
& G_{144} \\ \hline
\ea
\]

We must confess that unfortunately the direct application of our
technique in the previous section has yielded no favourable results,
i.e., no quotients groups of $G$ gave any of the exceptional $SU(3)$
subgroups $\Sigma_{36\times3, 72\times3, 216\times 3, 360\times3}$ or
nontrivial quotients thereof (and {\it vice versa}), even though the
fusion graphs for the former and the McKay quiver for the latter have
been pointed out to have certain similarities
\cite{DiFrancesco,Reflection,HanHe}. These similarities are a little
less direct than the Mckay Correspondence for $SU(2)$ and involve
truncation of the graphs, the above failure of a na\"{\i}ve
correspondence by quotients may be related to this complexity.

Therefore much work yet remains for us \cite{Prog}. Correspondences
for the infinite series in the $SU(2)$ case still needs be formulated
whereas a method of attack is still pending for $SU(3)$ (and
beyond). It is the main purpose of this short note to inform the
reader of an intriguing correspondence between WZW modular invariants
and finite groups which may hint at some deeper mechanism yet to be uncovered.

\section*{Acknowledgements}
{\small {\it Ad Catharinae Sanctae Alexandriae et Ad Majorem Dei
Gloriam...\\}}
We would like to thank J. S. Song for some initial collaboration and
discussions as well as A. Hanany for reviewing the manuscript. 
YHH would in particular like to extend his gratitude to
T. Gannon for many helpful insights as well as the organisers of the
``Modular Invariants, Operator Algebras and Quotient Singularities
Workshop'' and the Mathematics Research Institute of the University of
Warwick at Coventry, U. K. for providing the friendly atmosphere
wherein much fruitful discussions were engaged.

\end{document}